\documentclass[apj]{emulateapj}
\usepackage{graphicx,hyperref}
\usepackage{txfonts}
\usepackage{epstopdf}
\usepackage{longtable}
\usepackage{natbib}
\usepackage{color}
\usepackage{breakurl}

\newcommand{\scs}{\scriptsize}

\shorttitle{Molecules in IC~4997}
\shortauthors{N. K. Rao et al. }

\begin{document} 

\title{Detection of CH$^{+}$, CH  and H$_2$ molecules in the Young Planetary Nebula IC\,4997\thanks{Based on
observations obtained with The Nordic Optical Telescope and The Harlan J. Smith Telescope.}}

\author{N. Kameswara Rao}
\affil{Indian Institute of Astrophysics, Bangalore 560034, Karnataka, India}
\affil{The W. J. McDonald Observatory, Department of Astronomy, University of Texas at Austin, TX 78712, USA}
\email{nkrao@iiap.res.in (NKR)}

\author{David L. Lambert}
\affil{W.J. McDonald Observatory and Department of Astronomy, The University of Texas at Austin, Austin, TX 78712-1205, USA}
\email{dll@astro.as.utexas.edu (DLL)}

\author{Arumalla B. S. Reddy}
\affil{Indian Institute of Astrophysics, Bangalore 560034, Karnataka, India}
\email{balasudhakara.reddy@iiap.res.in (ABSR)}

\author{D.A. Garc\'{\i}a-Hern\'andez}
\affil{Instituto de Astrof\'{\i}sica de Canarias (IAC), E-38205 La Laguna, Tenerife, Spain}
\affil{Departamento de Astrof\'{\i}sica, Universidad de La Laguna (ULL), E-38206 La Laguna, Tenerife, Spain}
\email{agarcia@iac.es (DAGH)}

\author{Arturo Manchado}
\affil{Instituto de Astrof\'{\i}sica de Canarias (IAC), E-38205 La Laguna, Tenerife, Spain}
\affil{Departamento de Astrof\'{\i}sica, Universidad de La Laguna (ULL), E-38206 La Laguna, Tenerife, Spain}
\affil{Consejo Superior de Investigaciones Cient\'{\i}ficas, Madrid, Spain}
\email{amt@iac.es (AM)}

\author{J. J. D\'{\i}az-Luis}
\affil{Observatorio Astron\'{o}mico Nacional (OAN-IGN), Alfonso XII, 3 y 5, 28014  Madrid, Spain}
\email{jdiaz@iac.es (JJDL)}



\begin{abstract}
We have detected CH$^{+}$ and CH molecular absorption lines from  the young compact planetary nebula IC\,4997 from  high resolution optical spectra. A high-resolution infra-red (H and K bands) spectrum  provides detection of H$_2$ emission lines amongst many other lines. The H$_2$ lines provide an excitation temperature of 2100 K which may result from UV fluorescence in the envelope or from shocks formed at the interface between an expanding outflow of ionized gas and the neutral envelope ejected when the star was on the AGB. It is suggested that the CH$^+$ may result from the endothermic reaction C + H$_2$ $\rightarrow$ CH$^+$ + H. Intriguingly, CH$^{+}$ and also CH show a higher expansion velocity than H$_{\rm 2}$ emission suggesting they may be part of the post-shocked gas. 
\end{abstract}


\section{Introduction} 
IC\,4997 (PNG 058.3$-$10.9, PK\,058$-$10.1, IRAS\,20178$+$1634) is a young rapidly evolving compact bipolar planetary nebula (PN) discovered by Williamina Fleming -- see Pickering (1896). It  represents an  early phase of the transition from an AGB star to a canonical PN comprising ionized gas around a hot star approaching the top of the white dwarf cooling track. Objects in the earliest phases of the AGB to PN transition might be expected to retain a shell or disk of neutral gas ejected by the AGB star and which will soon be converted to ionized gas as the central star evolves to higher temperatures. This transition phase from neutral to ionized gas is little explored. In this paper, we examine IC\,4997's neutral envelope using high-resolution optical and infrared (H and K band) spectra which provide detections of optical lines of the CH and CH$^+$ radicals in absorption and of infrared H$_2$ quadrupole vibration-rotation transitions in emission. The molecules are part of the expanding neutral gas region around the central star.  (A thorough analysis of the many optical and infrared atomic emission lines and a detailed modelling of the nebula are not attempted here.)

Our detections of CH and  CH$^+$ are important because there are very few detection of CH$^{+}$ and CH optical lines from circumstellar environments and none from planetary nebulae. An historical account of discovery of absorption lines of CH$^{+}$ and CH from the interstellar and circumstellar medium is given by Oka et al. (2013). Only two detections of CH$^{+}$ (no lines of CH) in a circumstellar environment have been reported in the optical region:  emission  around the Red Rectangle nebula and  absorption around the post-AGB star HD 213985 (Bakker et al. 1997) and towards the young star Herschel 36 (Dahlstrom et al. 2013). Rotational transitions of CH$^{+}$ and CH have been seen in far-infrared spectrum of the planetary nebula  NGC 7027 (Cernicharo et al. 1997). Nine lines of the 1-0 vibration-rotation band of CH$^+$  near 3.6 $\mu$m  were detected from NGC 7027 by Neufeld et al. (2020).

IC\,4997 is classified as a young PN.  From  the ground  it appears almost stellar-like. An image from the {\it Hubble Space Telescope}  shows  a bright  compact central object  with faint bipolar lobes extending out about 1.4 arc seconds (Sahai, Morris \& Villar  2011).  Sahai  et al. give the distance as 2.3 kpc and the expansion age as 1825 yr.  Miranda, Torrelles \& Eiroa (1996) suggest the PN is as young as 675 yr. Miranda \& Torrelles (1998) from radio continuum observations at 3.6 cm and 2 cm report that they resolve the central compact object into an elliptical structure  within an extended equatorial disk which they suggest is the collimator for the bipolar lobes. Other radio observations have been reported by Miranda et al. (1996) and Gomez et al. (2002). For such a young PN, it is not surprising that several studies have reported evolution of the optical emission lines and high densities for the ionized regions. For example, Menzel, Aller \& Hebb (1941) noted  that the  intensity ratio of 4363 \AA\ [O\,{\sc iii}] to H$\gamma$  was exceptionally high.  Evolution of the emission lines was noted first by  Aller \& Liller (1966) and then among others by Kostyakova \& Arkhipova (2009) whose observations spanned forty years. From their modeling of the optical spectrum,  Hyung, Aller \& Feibelman  (1994) suggested that a dense inner shell must have been ejected at low velocity and gradually accelerated until it reached the presently observed expansion velocity of 16-20 km s$^{-1}$.
 
Young PNs such as IC\,4997 are expected to retain parts of the neutral gas ejected by the AGB. This neutral envelope or disk will be eroded eventually by the ionizing flux from the central star. Neutral gas around IC 4997 has  been revealed by the 21 cm hydrogen line in absorption (Altschuler et al. 1986)
and by the Na D line absorption at circumstellar velocities (Dinerstein, Sneden \& Uglum  1995). Perhaps surprisingly, neither CO  through its pure rotational emission lines at millimetre wavelengths (Zuckerman \& Gatley 1988; Huggins \& Healy 1989; Gussie \& Taylor 1995)  nor H$_2$ via its vibrational-rotational lines at 2 microns (Zuckerman \& Gatley 1988; Kastner et al 1996; Marquez-Lugo et al. 2015) have been detected in low resolution spectra. A puzzling observation discussed by Miranda et al. (1996) is that the H$\alpha$ emission line shows wings with a total width of nearly
5500 km s$^{-1}$.  Such wings are found exclusively for H$\alpha$. Lee \& Hyung (2000) proposed that the broad wings result from Raman scattering by atomic hydrogen. Our detections of CH, CH$^+$ and H$_2$ provide new probes of the neutral gas around IC\,4997 and also identifies a new supplier of CH$^+$ to the interstellar medium. 

Neutral gas, especially gas ejected by an AGB star, is expected to contain dust. Pottasch et al. (1984) and Lenzuni, Natta \& Panagia  (1989) noted that IC4997 has the very high dust to gas mass ratio of $\approx$ 0.02 . Mid-infrared spectra of IC 4997 obtained seem to show the presence of emission features in the 17 to 21 $\mu$m region and near 3.3 $\mu$m linked to PAH molecules found with dust grains (Garc\'{i}a-Hern\'{a}ndez -personal communication; Ohsawa et al 2016 ). 
  
This paper is organized as follows: the next section presents observations of the 2014 optical and 2016 near-infrared high-resolution spectra of IC 4997.  Section 3 presents an overview of the spectra. Section 4 discusses the status and properties of cool circumstellar envelope based on Na D profiles
including a comparison with lower resolution spectra from 1988-1989 by Dinerstein et al. (1995), the CH and CH$^+$ circumstellar lines, their possible modes of formation and explores their relationship to the diffuse interstellar bands and H$_{\rm 2}$ molecular emissions. Section 5 discusses the geometry of the neutral envelope and the ionized gas and the presence of H$_{\rm 2}$ and CH$^+$ and CH. The paper concludes by highlighting the results from our optical and infrared spectra concerning this young PN IC 4997.

\section{High-resolution spectra}

\subsection{Optical and Infrared Observations}
Observations of IC 4997 were obtained at the 2.5 m Nordic Optical Telescope (NOT) on La Palma (Observatorio del Roque de Los Muchachos, ORM) on 2014 July 6 with the fibre-fed FIES spectrograph. The spectra cover the interval 3600 -- 7200 \AA\ at a resolution of about 44000. Eight exposures of 1800 s each were obtained and, in the final combined IC 4997 spectrum, the S/N in the continuum ranges from about 43 at 4300 \AA\ to about 80 at 5900 \AA. The 1.3 second of arc fibre admitted essentially all the light from the star and nebula. The observed FIES spectra - processed with the FIES reduction software (FIEStool\footnote{See http://www.not.iac.es/instruments/fies/fiestool/ FIEStool-manual-1.0.pdf}) - were corrected for heliocentric motion and combined, and the stellar continuum was fitted by using standard astronomical tasks in IRAF\footnote{Image Reduction and Analysis Facility (IRAF) software is distributed by the National Optical Astronomy Observatories, which is operated by the Association of Universities for Research in Astronomy, Inc., under cooperative agreement with the National Science Foundation.} In interpreting the reduced spectrum, it is critical to recognize that strong emission lines bleed down the CCD rows and can appear as apparent emission lines in other orders.

Other high-resolution optical spectra are also considered. A spectrum was obtained on 2001 July 13 with the Utrecht echelle spectrograph (UES) at the 4.2 m William Herschel Telescope (ORM, La Palma). The wavelength coverage runs from 4300 \AA\ to 9000 \AA\ at a resolution comparable to that of our primary spectrum from the NOT.  However, the S/N ratio of the continuum is low but the red region of the UES spectrum is useful as a comparison with the 2014 NOT/FIES spectrum. Optical spectra were also obtained with the Tull spectrograph (Tull et al. 1995) at the W.J.  McDonald Observatory's 2.7 m telescope in 2016. These spectrum covered much of the optical spectrum at a resolution of about 60,000 with a good S/N ratio in the continuum and confirm the detections of molecular lines present in NOT/FIES.

High-resolution infrared spectra of IC\,4997 in the H and K bands were obtained on 2016 June 30 with the 2.7 m Harlan J. Smith telescope at the W.J. McDonald observatory using the IGRINS spectrograph (Park et al. 2014), which covers the whole H-band (i.e. 1.49 - 1.80 $\mu$m) and the K-band (i.e. 1.96 -- 2.46 $\mu$m) simultaneously at the resolving power of about 45000. The H- and K- band cameras use  2k $\times$ 2k HAWAII-2RG arrays as a detector.
Four individual exposures in both H- and K-bands of IC 4997 and a telluric standard were obtained in a sequence of two different positions on the slit separated by about 7 arcsec to subtract the sky contribution. The total exposure time was 600 s for each slit position. The spectroscopic data reductions were performed  using various routines available within the imred and echelle packages of {\scs IRAF}. Wavelength calibration was done using telluric absorption lines in the spectra of the fast-rotating A-type star observed at a very similar airmass. Telluric lines in the spectrum of IC 4997 in both H- and K-bands were removed by dividing the spectra of IC 4997 with that of the A-type star.

Our principal application of the IGRINS spectrum involves emission lines, specifically the fluxes of weak H$_2$ lines. Therefore, we attempted a flux
calibration of the spectrum. We rely on infrared photometry to estimate the continuum flux. Such a procedure assumes that the emission lines in these bands do not affect seriously the total continuum flux in the band.  IC 4997 is known as a variable in visual bands. Since simultaneous infrared photometry is not available for our observations,  we have to assess  how much of a variable it is in infrared bands and  whether there is any secular variation in the infrared photometry. Fortunately, IC\,4997 appears to be only slightly reddened.

Taranova \& Shenavrin (2007) monitored IC\,4997 at J, H, K and L from 1999 October 3 to 2006 August 9. In addition, the 2MASS catalogue provides observations on 1998 September 26. These observations suggest that the infrared magnitudes were pretty much constant from 1998 to 2006. Therefore, we  use an average measurement of H and K.  The interstellar reddening has been estimated as E(B-V) = 0.27  (Flower 1980) and 0.19 (Phillips \& Cuesta 1994). We adopt E(B-V) =0.23 which leads to  extinction of A(H) =0.134  and A(K) =0.074 from the standard relations for the ISM (Cardelli, Clayton \& Mathis 1989).  Then, we obtain H = 10.5 and  K = 9.8  and adopt the standard flux calibration (Tokunaga 2000).  Emission line fluxes are measured with respect to the flux-calibrated continuum.

Several checks against published line fluxes suggest our calibration is satisfactory.  For example, the  flux of Brackett $\gamma$ estimated from our  IGRINS spectrum is 24 $\times$ 10$^{-13}$ erg cm$^{-2}$ s$^{-1}$. Sterling \& Dinerstein (2008) observed IC 4997 at a resolving power of 500 on 2004 November 7 and their flux calibration gave the Brackett $\gamma$ flux as 19.2$\pm$0.4$\times$10$^{-13}$ erg cm$^{-2}$ s$^{-1}$. Geballe, Burton \& Isaacman (1991) observed IC 4997 on 1990 November 30 at a resolving power of 250 and reported a Brackett $\gamma$ flux of 28$\times$10$^{-13}$ erg cm$^{-2}$ s$^{-1}$. These estimates bracket our measurement.  Sterling \& Dinerstein (2008) also provide flux of [Fe\,{\sc iii}] 2.218$\mu$m line  as 1.36 $\times$ 10$^{-14}$ erg cm$^{-2}$ s$^{-1}$  We obtain the flux of this line as 1.41 $\times$ 10$^{-14}$ erg cm$^{-2}$ s$^{-1}$  in remarkable agreement with Sterling \& Dinerstein's value. Marquez-Lugo et al.'s (2015) estimate of the Brackett $\gamma$ flux is 6.1 $\times$ 10$^{-13}$ erg cm$^{-2}$ s$^{-1}$, which is inconsistent with the other estimates above unless the line fluxes are highly variable. However, their observations were obtained during July 2004, about the same time as Sterling \& Dinerstein's observations; thus we conclude that the flux calibration of Marquez-Lugo et al. (2015) may be uncertain.

\section{Overview of the  spectra}

\subsection{The optical spectrum}
The optical spectrum is dominated by nebular emission lines with a wide range in excitation and ionization: a list of lines is provided and discussed by Hyung et al. (1994).  The few absorption lines associated with the PN include the Na D lines previously discussed by Dinerstein et al. (1995) and our detections of CH and CH$^+$ lines. Discussion of the emission and absorption line profiles is aided by reference to several key velocities. First, velocities associated with the star and nebula are well separated from those tied to the intervening interstellar medium. Nebular absorption lines are shown in Figure 1. Strong interstellar components at Na D and Ca\,{\sc ii} K are   at a heliocentric velocity of 0 km s$^{-1}$ flanked by weaker components at $-21$ and $+15$ and $+23$ km s$^{-1}$. Nebular Na D absorption well separated from interstellar components  appears at $-84$ and $-97$ km s$^{-1}$ with broad emission  peaking at about $-66$ km s$^{-1}$. The CH and CH$^+$ lines coincide with the $-97$ km s$^{-1}$  nebular Na D component.  Nebular absorption is not seen at the Ca\,{\sc ii} K line.  In addition to the nebular  absorption lines, a few absorption lines arise from the hot central star. Weidmann, M\'{e}ndez \& Gamen  (2015) reported stellar absorption lines  of He\,{\sc ii} at 4200, 4512 and 5412 \AA. In our spectrum, the 5412 \AA\ line is blended with [Fe\,{\sc iii}] 5411.98 \AA\ in emission  but the other two He\,{\sc ii} lines are present at a mean radial velocity of $-116$ km s$^{-1}$.

Many nebular emission lines appear as two partially resolved lines with a red component close to $-51$ km s$^{-1}$ and the blue component with  a peak velocity near $-79$ km s$^{-1}$. The mean velocity of $-65$ km s$^{-1}$ might be considered to be the systemic velocity with 14 km s$^{-1}$  as the expansion velocity - see below.  Figure 2 shows several profiles of forbidden emission lines with their characteristic double-peaked profiles. In general, the blue peak of such emission lines is stronger than the red peak. In some cases, the blue to red asymmetry is such that the line may appear almost as a single line shifted to the blue of the systemic velocity: the [O\,{\sc iii}] and Balmer lines are examples of this kind of profile.
Permitted lines such as the Si\,{\sc ii} 6371 \AA\ line typically show single-peaked emission with an asymmetry suggesting that emission to the blue of the systemic velocity is stronger than the emission to the red. Changes in the relative intensities of emission lines have long been recognized and documented (see, for example, Kostyakova \& Arkhipova 2009). Contributions of our 2014  and 2016 spectra to this topic may be considered elsewhere.

\begin{figure*}
\centering
\includegraphics[width=16cm,height=15cm]{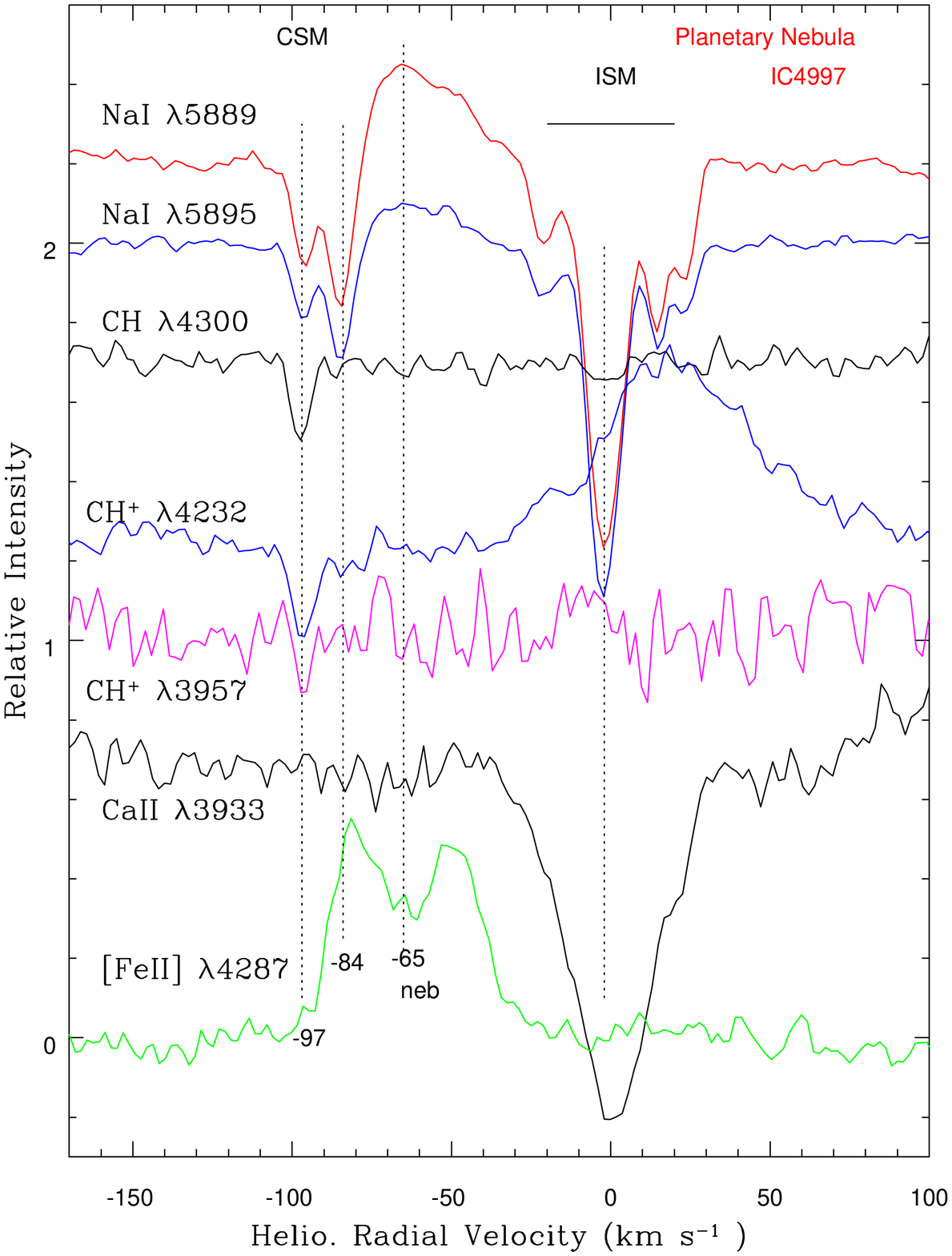}
\vspace{-0.3cm}
\caption{Absorption lines towards the central star of IC\,4997 obtained on 2014 July. The interstellar medium (ISM) and nebular lines (CSM) are clearly separated. Although neutral atomic  and molecular lines have nebular components, the  Ca\,{\sc ii} K line, though very strong in the ISM, does not have any nebular absorption component. Nebular [Fe\,{\sc ii}] emission line is plotted to show the radial velocity of emission components and the systemic velocity. The dashed vertical lines refer to the radial velocity of various components. The relative intensity scale (1.0 to 0.0) is the same for all lines but each line's continuum level is displaced by an arbitrary amount. }
\end{figure*}

\begin{figure*}
\centering
\includegraphics[width=16cm,height=15cm]{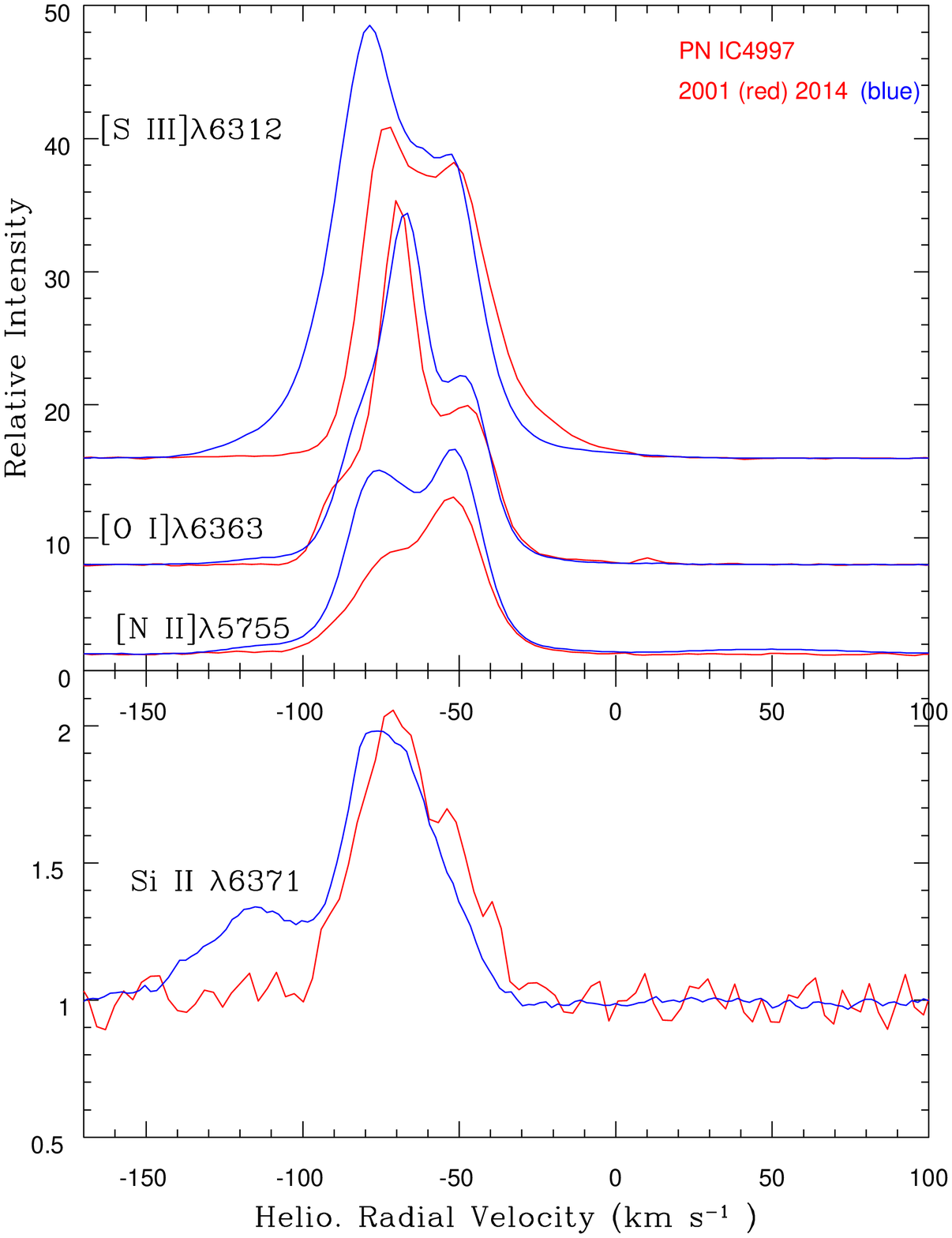}
\vspace{-1.1cm}
\caption{Emission lines  of IC\, 4997 during 2001 July (red) and 2014 July (blue). Lines of various levels of excitation are shown.  Although some high excitation lines show enhanced blue components in 2014 relative to 2001 spectrum, no major changes are seen. The bump shortward of the Si\,{\sc ii} 6371 \AA\ emission line in the 2014 spectrum is an artifact arising from charge transfer from the strong [O\,{\sc i}] 6300 \AA\ in the next order.}
\end{figure*}

\subsection{Overview of the near-infrared spectrum}
The spectrum consists of emission lines of H\,{\sc i}, He\,{\sc i}, Mg\,{\sc ii}, [Fe\,{\sc iii}], [Fe\,{\sc ii}] , H$_{\rm 2}$  and some unidentified lines. Selected lines are shown in Figure 3. The dominant lines are from the H\,{\sc i} Pfund and Brackett series.  These lines are symmetrical and well represented by Gaussian profiles. The profiles  He\,{\sc i} lines are slightly asymmetric with a steeper rise on the blue side and slower decline on the  red side. The Mg\,{\sc ii} , [Fe\,{\sc iii}] and particularly the [Fe\,{\sc ii}] lines appear as a blend of two components. The H$_{\rm 2}$ lines appear single.

A selection of the  He\,{\sc i} , H\,{\sc i} , Mg\,{\sc ii}, [Fe\,{\sc iii}], [Fe\,{\sc ii} ] and H$_{\rm 2}$ lines were measured  for radial velocity by fitting Gaussian  profiles. Fourteen He\,{\sc i} lines give a mean heliocentric radial velocity of $-$72.4$\pm$1.6 km s$^{-1}$. Thirty one H\,{\sc i}  lines  provide a similar mean radial velocity of -71.3$\pm$1.9 km s$^{-1}$. (Although the He\,{\sc i} lines  are asymmetric, the centre of the profile has been measured.) [Fe II], [Fe III],  and Mg II lines are centered at the heliocentric radial velocity of about $-65$ km s$^{-1}$  with the red and blue components displaced approximately symmetrically with respect to the central velocity: the mean velocity of the red component of the [Fe\,{\sc ii}] and [Fe\,{\sc iii}] lines is at $-77\pm2$ km s$^{-1}$ and the blue component is at $-50\pm2$ km s$^{-1}$.  The infrared profiles match the optical profiles when the same ions are compared. In marked contrast, as Figure 3 clearly shows, the H$_{\rm 2}$ emission presents a heliocentric radial velocity of $-85.9\pm1.9$ km s$^{-1}$ which is at the blue limit of the low-excitation lines of the Mg and Fe ions.  The two strongest  H$_2$ lines are weakly in emission also at $-27$ km s$^{-1}$ which corresponds to the red limit of the Mg and Fe lines. It appears that the H$_2$ profiles have an intimate relationship to the extremes of the region emitting the lines from the Mg and Fe ions.

\begin{figure*}
\centering
\includegraphics[width=17cm,height=18cm]{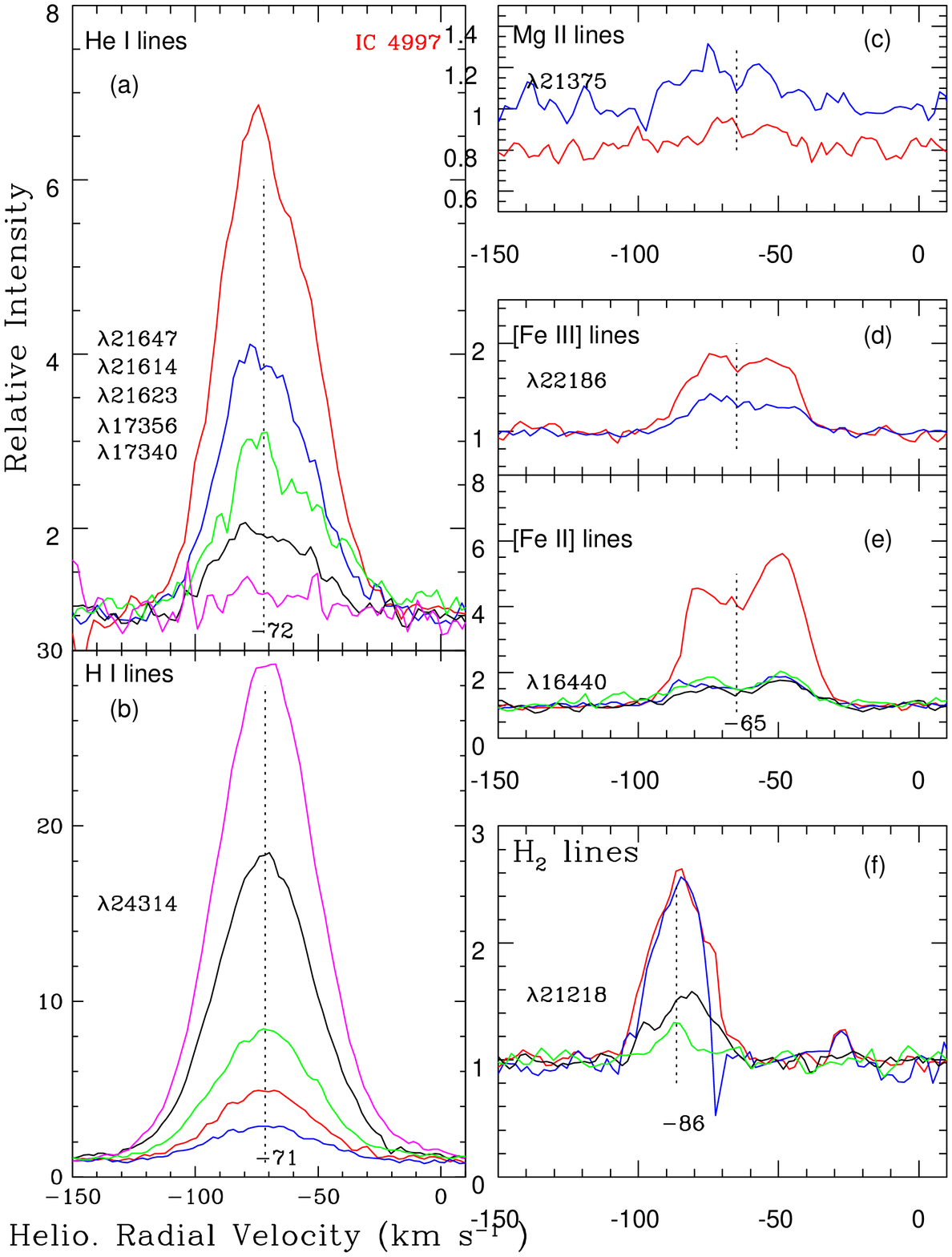}
\vspace{-1.5cm}
\caption{ The profiles of various infrared lines-(a) He\,{\sc i} lines :  $\lambda$21647.389 - red, $\lambda$21613.694 - blue, $\lambda$21622.906 - black, $\lambda$ 17356.452 - green, $\lambda$17340.448 - magenta; (b) H\,{\sc i}  lines :$\lambda$24313.620 Pf20 - red, $\lambda$23743.770 Pf25 - blue, $\lambda$15884.897 Br14 - black, $\lambda$16811.10 Br11 - black, $\lambda$15345.999 Br18 - green; (c) Mg\,{\sc ii} lines: $\lambda$ 21437.96  - red,
 $\lambda$ 21374.75  - blue ; (d) [Fe\,{\sc iii}] : $\lambda$ 22186.7 - red, $\lambda$21457.3 - blue; (e) [Fe\,{\sc ii}] : $\lambda$16440.017 - red, $\lambda$16773.406 - blue, $\lambda$15999.146 - black, $\lambda$ 15338.939 - green; (f)H$_{\rm 2}$  lines : $\lambda$21218.334 1-0 S(1)- red, $\lambda$ 24065.914 1-0 Q(1)- blue, $\lambda$ 20337.576 1-0 S(2)- black, $\lambda$ 22477.173 2-1 S(1) - green.}
\end{figure*}

\section{The cool circumstellar envelope}
    
\subsection { Nebular Na D and Ca\,{\sc ii} K }
The Na D profiles are shown in Figure 4 and compared in Figure 1 with other probes of the cool circumstellar envelope. A complex of interstellar Na D
absorption components is present with the strongest line at a heliocentric velocity of  $-1$ km s$^{-1}$ flanked by weaker components to the red and blue. Contributions in absorption  from the planetary nebula  at negative radial velocities were first reported by Dinerstein et al. (1995) from spectra at a resolution of 10 km s$^{-1}$ taken in 1988 and 1989. Our higher resolution spectra from 2014  and 2016  reveal structure not clearly seen in Dinerstein et al.'s  spectrum. Nebular Na D emission is present with peak intensity near the systemic velocity of $-66$ km s$^{-1}$ with a half width of about 15$\pm$1 km s$^{-1}$ which velocity may be the expansion velocity of the Na-containing shell. Nebular Na D absorption is present with overlapping components at $-84$ and $-97$ km s$^{-1}$ with the former slightly stronger than the latter. The $-84$ km s$^{-1}$ component is almost coincident with the blue peak of  doubly-peaked emission lines  and emission peak of H$_2$ lines (see Figures 2 and 3). The $-97$ km s$^{-1}$ absorption component has no obvious counterpart among the nebular emission lines and their components but it coincides well with the CH and CH$^+$ absorption (Figure 1).

Nebular absorption at $-83$ km s$^{-1}$ in the Na D lines with a hint of emission was reported by Dinerstein et al. They did not resolve the absorption components in either the interstellar or the circumstellar lines. The Na D absorption appears to have weakened considerably between 1988-1989 and 2014. For 2014, the equivalent width of the total absorption measured by Gaussian deconvolution is 119 m\AA\ for D$_2$ and 85 m\AA\ for D$_1$ but for the
1988-1989 spectra Dinerstein et al. reported 270$\pm$15 m\AA\ and 275$\pm$30 m\AA\ for D$_2$ and D$_1$, respectively. This is a striking weakening of the circumstellar D lines. (Equivalent width measurements of the interstellar lines are in good agreement: the (D$_2$, D$_1$) equivalent widths in m\AA\  are (413, 292) in 2014 and (400, 300) in 1988-1989). For the circumstellar Na\,{\sc i} D lines, whose equivalent widths show that they are not optically thin, we used the doublet ratio method (Spitzer 1968) to obtain column densities: $= 9.4 \times10^{11}$ cm$^{-2}$ for the $-84$ km s$^{-1}$  and $= 4.7 \times 10^{11}$ cm$^{-2}$ for the $-97$ km s$^{-1}$ component.   Circumstellar Na D  weakened markedly over the last three decades. Emission at Na D is roughly centred at the systemic velocity  which implies that Na atoms are approximately uniformly distributed about the central star. 
 
A noticeable feature of the circumstellar absorption is the absence of a Ca\,{\sc ii} K line. Figure 1 shows that the interstellar Na D and Ca\,{\sc ii} K line  are of comparable strength but the K line is not detectable at circumstellar  velocities. A possible interpretation is that Ca is very highly depleted into and onto grains in IC\,4997’s very dusty neutral envelope. As is well known, calcium is severely depleted in the diffuse ISM but the Ca depletion in this PN’s neutral envelope must be extremely severe. Photoionization by the hot central star may also play a role but, if Ca and Ca$^+$ are greatly photoionized, one expects severe loss of molecules CH and CH$^+$ as well as Na atoms.

\begin{figure*}
\centering
\includegraphics[width=16cm,height=12cm]{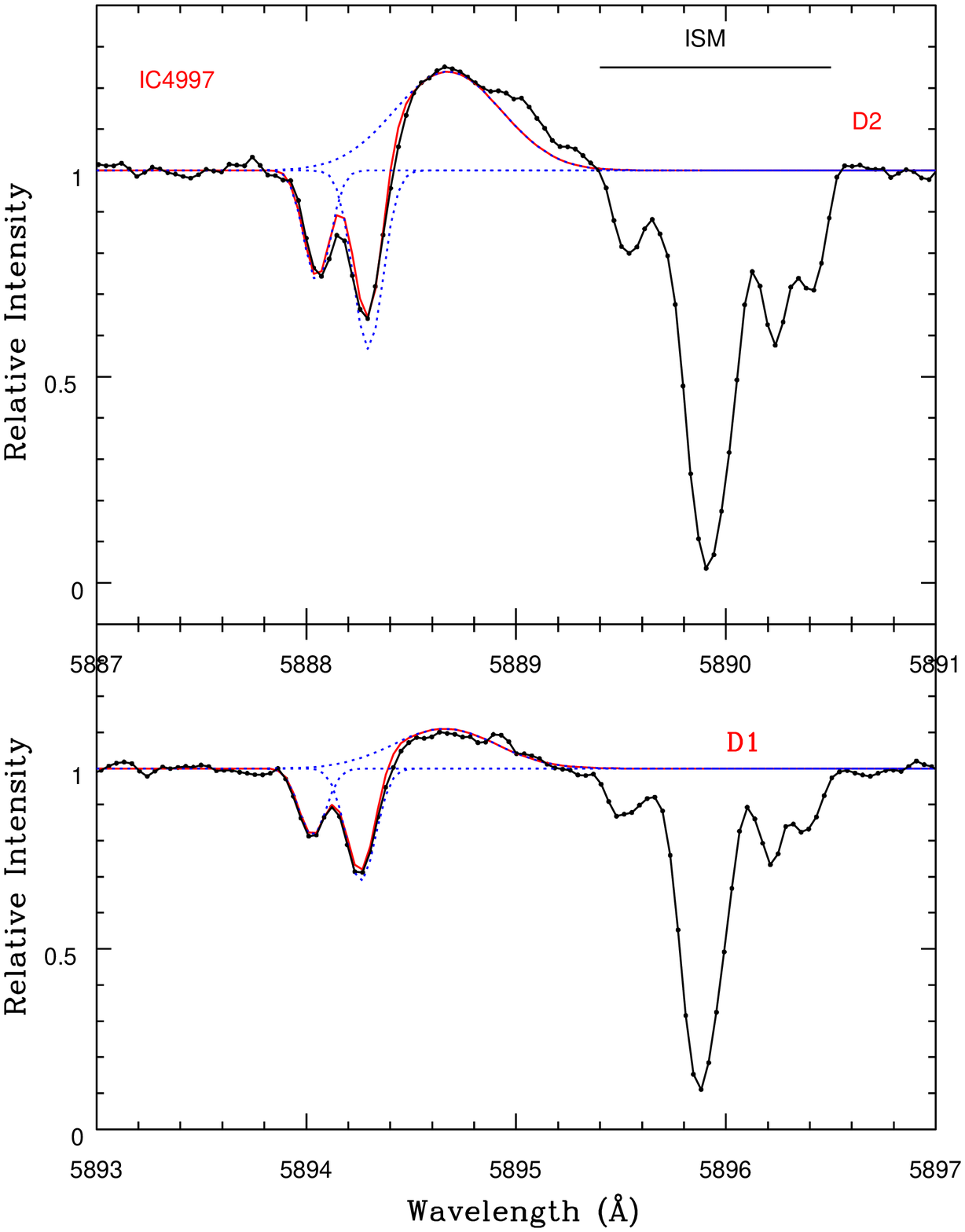}
\vspace{-0.7cm}
\caption{Na\,{\sc i} D  lines towards the central star of IC\,4997 from spectra obtained on 2014 July. The interstellar medium (ISM) and nebular lines (CSM) are clearly separated. Na\,{\sc i} D line emission  is prominent and centred on the nebular velocity. The nebular profiles  are decomposed  into three Gaussian components (blue dotted profiles) consisting of an emission and two absorption component. The combined profile (red) is matches  the observed profile (black). }
\end{figure*}

\subsection{CH and CH$^{+}$}
A surprising discovery in the NOT/FIES spectrum was the presence in absorption of  a single transition of both CH and of CH$^+$ (Figure 1). Both lines
originate from their respective molecule's lowest bound level. Detected lines appear at the heliocentric velocity of $-97$ km s$^{-1}$ which is coincident with  the bluemost of the two Na D nebular  absorption components. There is no detectable molecular absorption at $-84$ km s$^{-1}$, the velocity of the redder of the two Na D circumstellar absorption lines. Also, the $-97$ km s$^{-1}$  absorption is outside the velocity range spanned by profiles of weak permitted and forbidden emission lines. Presence of these molecular lines is confirmed by the 2016 McDonald spectra.

The CH line is the R$_2$(1/2) doublet of the 0-0 band of the A-X electronic  transition, a line from the lowest bound level and one seen commonly along lines of sight through the diffuse interstellar medium. The rest wavelength is 4300.313 \AA\ and the absorption $f$-value is $f = 0.00506$ (Lambert \& Danks 1986; Gredel, van Dishoeck \& Black  1993). Lines from the first few excited levels were searched for and not found. The corresponding lines from the 0-0 band of the B-X transition at 3890.217 \AA\, 3886.409 \AA\ and 3878.774 \AA\ were not found but upper limits to their equivalent widths yield upper limits to the CH column density consistent with that obtained from the equivalent width of 21 m\AA\ for the 4300 \AA\ line on the assumption that the line is unsaturated: N(CH) $= 2.5 \times 10^{13}$ cm$^{-2}$.

The CH$^{+}$ line is the R(0) line from the 0-0 band of the A-X electronic transition. The laboratory wavelength is 4232.548 \AA\ and the $f$-value is 0.00545. The corresponding line from the transition's 1-0 band is at 3957.692 \AA\ with $f$ = 0.00331. Although the line is not clearly detected, the upper limit on the column density is consistent with that determined from the 34 m\AA\ equivalent width of the 4232 \AA\ line: N(CH$^{+}$)$ = 3.9 \times$ 10$^{13}$ cm$^{-2}$. The widths (FWHM) of the CH$^{+}$ line at $9\pm1.5$ km s$^{-1}$ and the CH line at 7$\pm1.5$ km s$^{-1}$ are equal to within the measurement uncertainties.

The chemistry providing CH and CH$^{+}$ in the $-97$ km s$^{-1}$ circumstellar gas may not be very different from that in the diffuse interstellar medium. Surveys of CH (Danks, Federman \& Lambert 1984) and CH$^{+}$ (Lambert \& Danks 1986) in the ISM provide nine lines of sight with both radicals
present with a mean ratio of the equivalent widths of CH$^{+}$ and CH of 1.7 and a small spread from 1.1 to 2.9.  IC 4997 provides a ratio of 1.6, a value close to the mean for the ISM. 
In the case of the ISM, Lambert \& Danks noted the CH$^+$ column density was correlated with the measured column density of rotationally excited H$_2$ molecules and suggested that CH$^+$ was produced by the endothermic reaction C + H$_2$  $\rightarrow$ CH$^+$ + H in warm gas.

One other molecule featured in absorption in the ISM is CN with the Violet B-X 0-0 band providing the strongest line R(0) at 3874.608 \AA\ and weaker lines R(1) at 3873.998 \AA\ and P(1) at 3875.763 \AA. These are not detectable in our spectrum.  The equivalent width limit of 4 m\AA\ for the R(0) line corresponds to a column density limit of 9 $\times 10^{11}$ cm$^{-2}$.  For an estimate of the CN to CH column density in the ISM, we refer to Federman, Danks \& Lambert (1984) whose Table 2 gives column densities for seven lines of sight. With the exception of one outlier (HD 29647), the mean  CN to CH column density ratio is 0.13 with a spread from 0.05 to 0.22 but 1.1 for the outlier. With the mean ratio and the CH column density, the expected CN column density is  3 $\times$ $10^{12}$ cm$^{-2}$ or three times greater than the estimated upper limit. This discrepancy is not too surprising considering that the chemistry of CN formation and destruction is surely different in  hot circumstellar and the cold diffuse cold ISM.
  
Recently, Oka et al. (2013) established a connection between CH$^{+}$ and CH absorption lines from excited levels and the shape and strength of the diffuse interstellar bands (DIBs) $\lambda$$\lambda$5780, 5797 and 6613 with the implication that the DIBs may arise from gas phase molecules. Oka et al.'s suggestion led us to search  for diffuse interstellar bands (DIBs) at circumstellar  velocities. Some DIBs are obscured by emission features.  Rest wavelengths of the DIBs were taken from Hobbs et al. (2008). DIBs at 5780, 6376 and 6613 \AA\ are seen at the velocities of the principal interstellar lines (Na D and Ca\,{\sc ii} K) but not at circumstellar velocities.  This is probably not surprising because DIB strength in the ISM  roughly correlates with Na D strength and the circumstellar Na D absorption is much weaker than the ISM features.               

A noticeable difference between the circumstellar CH and CH$^+$ lines and their Na D counterpart is the absence of  emission accompanying the CH and CH$^+$ lines. Photons absorbed by Na atoms in the resonance D lines are reemitted in these lines. In the case of CH (and also CH$^+$), a great majority of the photons absorbed in the observed line will be reemitted in the same line and, thus, the lack of an emission component implies that the CH and CH$^+$ molecules are concentrated along the line of sight to the star whereas the Na atoms are approximately distributed symmetrically about the star with many atoms sensed only by the emission they provide. The CH and CH$^+$ molecules based on radial velocity appear to be co-located with the $-97$ km s$^{-1}$ Na D atoms. It does not seem possible to determine  if, as it should, this Na D component lacks an emission component.

\begin{table*}
\centering
\begin{minipage}{120mm}
\caption{Near-infrared Emission Lines of H$_{\rm 2}$ in IC 4997.  }
\label{default}
\begin{tabular}{cccccccc}  \hline
$\lambda\,$(vac)$^{a}$& Species & Transition &$E_{\rm upper}$$^{b}$  &$v_{\rm \odot}$&Eq.w&$F_{\rm \lambda}$$^{c}$&FWHM$^{d}$   \\
  \AA          &        &            &  K &km s$^{-1}$ &\AA & 10$^{-15}$ &km s$^{-1}$   \\  \hline
24237.312 & $H_{\rm 2}$   &   1-0 Q(3) & 6956 &-89.9  &1.43& 7.30   &17.6   \\
24065.941 & $H_{\rm 2}$   &   1-0 Q(1) & 6149 &-85.3  &2.44&12.50   &16.2    \\
22477.214 & $H_{\rm 2}$   &   2-1 S(1) &12550 &-87.3  &0.32& 1.71   &13.1   \\
22232.993 & $H_{\rm 2}$   &   1-0 S(0) & 6471 &-88.3  &0.74& 3.94  &20.3    \\
21542.253 & $H_{\rm 2}$   &   2-1 S(2) &13150 &-87.4  &0.11& 0.62   &(5.0)   \\
21218.313 & $H_{\rm 2}$   &   1-0 S(1) & 6956 &-86.3  &2.76&15.94   &22.3    \\
20337.563 & $H_{\rm 2}$   &   1-0 S(2) & 7584 &-84.0  &0.94& 5.88   &23.0    \\   \hline
\end{tabular}
\flushleft{
$^a$: Wavelengths calculated from energy levels computed by Komasa et al. (2011) \\
$^b$: Upper energy level $E_{\rm upper}$ and A-values are from Geballe (see text).\\
$^c$: in units of erg cm$^{-2}$ s$^{-1}$\\
$^d$:  Gaussian full width at half maximum corrected for instrumental broadening. 
           }
\end{minipage}
\end{table*}

\begin{table*}
\centering
\begin{minipage}{120mm}
\caption{Comparison of observed H$_{\rm 2}$ line flux ratios with model predictions.}
\label{tab:lineratmodelcomp}
\begin{tabular}{ccccp{0.03cm}ccccp{0.03cm}cc} \hline
Wavelength & Line ID  & \multicolumn{2}{c}{Observed flux ratios} & & \multicolumn{6}{r}{Line flux ratios from models}  \\  \cline{3-4} \cline{6-12}
$\lambda$ (\AA) &  &\multicolumn{1}{c}{IC4997} &\multicolumn{1}{c}{T-Tauri} & & \multicolumn{4}{c}{Sternberg \& Dalgarno (1989)} & & \multicolumn{2}{c}{Shocks$^{a}$} \\  \cline{3-4} \cline{6-9} \cline{11-12}
          &        &        &      & &  2A   &  2B  &  2C    &  2D & & C35 & J15  \\   \hline
24237.289 & 1-0 Q(3) & 0.46 & 0.77 & &  0.70 & 0.70 & 0.70   &0.70 & &0.70 &0.70  \\
24065.914 & 1-0 Q(1) & 0.78 & 0.74 & &  0.86 &$\leq$0.19&1.86&1.12 & &0.77 &0.63  \\
22477.173 & 2-1 S(1) & 0.11 & 0.09 & &  0.56 & 0.75 &$\leq$0.07&0.02& &0.05&0.24  \\
22232.898 & 1-0 S(0) & 0.25 & 0.23 & &  0.48 & $\ldots$ & 0.34   &0.29& &0.22 &0.20 \\
21542.118 & 2-1 S(2) & 0.04 & 0.04 & &  0.32 & $\ldots$ &$\leq$0.07&$\leq$0.02& &0.02&0.10 \\
21218.334 & 1-0 S(1) & 1.00 & 1.00 & &  1.00 & 1.00 & 1.00   &1.00 &  &1.00 &1.00  \\
20337.576 & 1-0 S(2) & 0.37 & 0.21 & &  0.61 & 0.79 & 0.17   &0.28 &  &0.35 &0.41  \\  \hline
\end{tabular} \\
\begin{footnotesize}
$^{a}$ : Smith (1995).   \\
\end{footnotesize}
\end{minipage}
\end{table*}

\subsection{H$_{\rm 2}$ emission lines in IC\,4997 }
Seven quadrupole vibration-rotation lines of H$_2$ were identified in the K band spectrum - see Table  1 and Figure 3.   Previous detections from low-resolution spectra of H$_2$ emission from IC\,4997 are limited to Geballe et al.'s (1991) measurement of the 1-0 S(1) line and Marquez-Lugo et al.'s (2015) claim of detections of 1-0 S(1), 2-1 S(1) and 8-6 S(3). Their reported flux of the 1-0 S(1) line is in good agreement with our measurement but their flux for the 2-1 S(1) line is a factor of about six greater than our measurement.

The Boltzmann plot corresponding to the measured line fluxes  is shown in Figure  5.  Energy levels and  Einstein A-values have been taken from Geballe
(\url{http://www.gemini.edu/sciops/instruments/nir/wavecal/h2lines.dat}). The slope of the best-fitting straight line gives an excitation temperature of 2100 K. The one discrepant point in Figure 5 comes from the 1-0 Q(3) line which fell in a region of low S/N and is subject also to correction for telluric H$_2$O lines. An indication that the measured flux of this line was likely underestimated is provided by the flux ratio of the two emission lines 1-0 Q(3) and 1-0 S(1) arising from the same upper level with their Einstein A-value ratio predicting a flux ratio of 0.70. The observed ratio is 0.46 and not the 0.70 of the A-values. (A correction for  (severe) reddening can only increase the ratio above 0.70.)

Discussion  of H$_2$ emission lines focusses on possible excitation mechanisms and not on the formation of the molecules. Our assumption is that formation and excitation of H$_2$ molecules are de-coupled processes and the molecules were  in the dusty neutral envelope prior to excitation by ultraviolet radiation from the hot central star and by shocks. Before considering the excitation on H$_2$ in IC 4997, we note that the flux ratios of lines in Table 1 are similar to those reported by Beck et al. (2008) for H$_2$ lines from a sample of five T Tauri stars where the relative importance of ultraviolet radiation and shocks to  H$_2$ excitation surely differs from their roles in IC 4997. Table 2 shows that the line ratios for the T Tauri stars are in good agreement with those for IC 4997. Such a consistency from two very different circumstellar environments suggests that separating contributors to H$_2$ excitation may not be a trivial exercise. This situation was well appreciated by Sternberg \& Dalgarno (1989) and appreciated by Beck et al. (2008). Sternberg \& Dalgarno  who in their concluding section of a study of molecular gas exposed to ultraviolet radiation ``Quiescent dense neutral gas, which is heated radiatively by UV photons, emits thermal H$_2$ line radiation that may, in practice, be difficult to distinguish from similar line emission produced in shock-heated gas." Beck et al. recognized the likely difficulty in writing ``it is not possible for us to distinguish conclusively between UV Ly$\alpha$ pumping of high-density gas and shock excitation as the main H$_2$ stimulation mechanisms in the inner regions of the stars." Distinction may be advanced by considering additional information such as line profiles.

Aware of the likely ambiguity concerning excitation mechanisms, we confront the relative line fluxes $F_\lambda$ in Table 1 with predictions for H$_2$ excitation under UV radiation and for excitation in gas heated by shocks. In all examples drawn from the literature, the predicted line ratios depend on various assumptions adopted in the modeling or, equivalently, the illustrated predictions are subject to adjustment by changing one or more parameters, that is exact reproduction of observed line ratios is not to be expected. An overall goal is to account for the excitation temperature of around 2000 K. Furthermore, we ignore the possibility that H$_2$ excitation may carry an imprint of the mechanisms by which H$_2$ is formed.

In the case of molecular gas exposed to UV radiation, line ratios are computed from line fluxes predicted by Sternberg \& Dalgarno (1989)  for four models of increasing total density $n_T$ exposed to the same UV flux which is 100 times the ambient interstellar flux. Sternberg \& Dalgarno explain why the excitation temperature increases with increasing $n_T$. At $n_T < 10^4$ cm$^{-3}$, the temperature is about 100 K and reaches about 1000 K at $n_T \sim 10^5$ cm$^{-3}$. The line ratio 1-0 S(1) to 2-1 S(1) is a simple indicator of the temperature.  Predicted  line ratios for Sternberg \& Dalgarno's (see their Table 2)  models 2A, 2B, 2C and 2D corresponding to $n_T = 10^3, 10^4, 10^5$ and $10^6$ cm$^{-3}$ are given in Table 2.  There is a fair correspondence between models for the higher densities with the observed line ratios. Low density  models 2A and 2B fail to account for the 2-1 S(1) and S(2) lines (relative to 1-0 S(1)). Sternberg \& Dalgarno note that at the lower densities their predictions are in excellent agreement with the earlier calculations of UV excited H$_2$ by Black \& Dalgarno (1987) whose calculations were done for low densities and, therefore, predicted low excitation temperatures.

Predictions for H$_2$ gas subjected to shocks are taken from Smith (1995 - see also Smith 1994) for a C-type shock at 35 km s$^{-1}$ and a J-type shock at 15 km s$^{-1}$. Continuous or C-type shocks show continuous magnetohydrodynamic parameters across the shock front. J-type shocks are ones in which the magnetohydrodynamic parameters are discontinuous (`jump`) across the shock front. Predictions depend on several parameters entering into the calculations and refinement of predictions to fit a set of observed line ratios is possible. Smith's predictions for his C35 and  J15 shocks are given in Table 2.  It will be noted that the C35 shock underestimates the relative  strengths of the 2-1 S(1) and S(2) lines -- the most excited lines in Table 1 -- and the J15 shock overestimates this pair's relative strengths. Inspection of Smith (1994) suggests that the two sets of predictions may be brought closer to the observations by increasing the velocity of the C-type shock and decreasing the velocity of the J-type shock but in neither case may a close fit be achieved by a velocity adjustment alone and other ingredients will deserve scrutiny.

In brief, our analysis of the H$_2$ lines detected from IC 4997 is unable to determine whether the dominant excitation mechanism is either UV fluorescence at a density sufficient for collisional redistribution of populations among the vibrational and rotational levels of H$_2$  or excitation of H$2$ resulting from a C-type or J-type shock front. In part, our inability results from the multi-parameter nature of the theoretical processes under study compounded by the limited array of published predictions for  H$_2$ line fluxes.  But these competing processes can both result, as here with IC 4997, in similar thermalized distributions of populations among rotational and low vibrational levels and effectively erasing the signatures of different excitation mechanisms. This was recognized by our quotation from Sternberg \& Dalgarno (1989) and its echo by Beck et al. (2008). This ambiguity calls for a detailed study of H$_2$ emission from planetary nebulae  such as spatially resolved mapping of the emission lines, investigations of  H$_2$  emission line profiles, etc.

\begin{figure*}
\centering
\includegraphics[width=14cm,height=14cm]{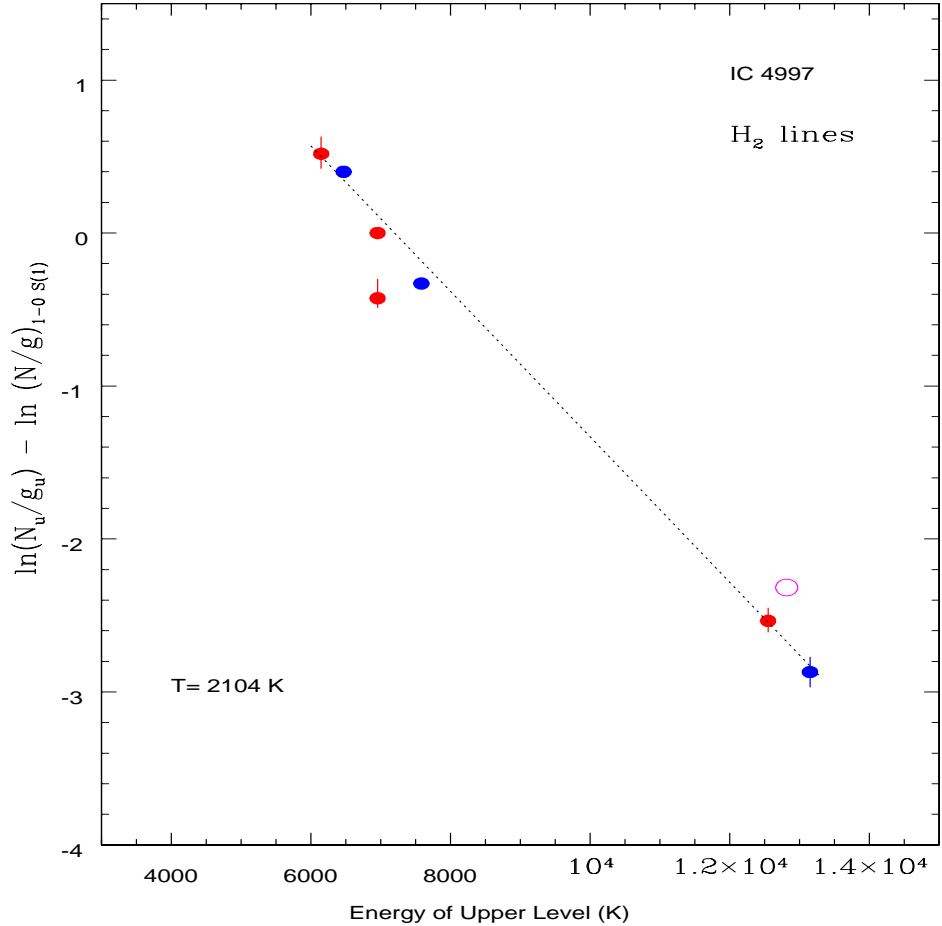}
\vspace{-1.0cm}
\caption{ The Boltzmann plot of H$_{\rm 2}$ lines in IC\,4997. The red dots refer to the  lines
 from ortho-H$_2$ (odd J) and blue dots to para-H$_2$ (even J) states. The circle in magenta 
refers to the upper limit of the 1-0 S(7) line. The errors in equivalent widths are shown as vertical
 lines around the value. For some lines the error is within the dot.  }
\end{figure*}

\begin{figure*}
\centering
\includegraphics[width=15cm,height=13cm]{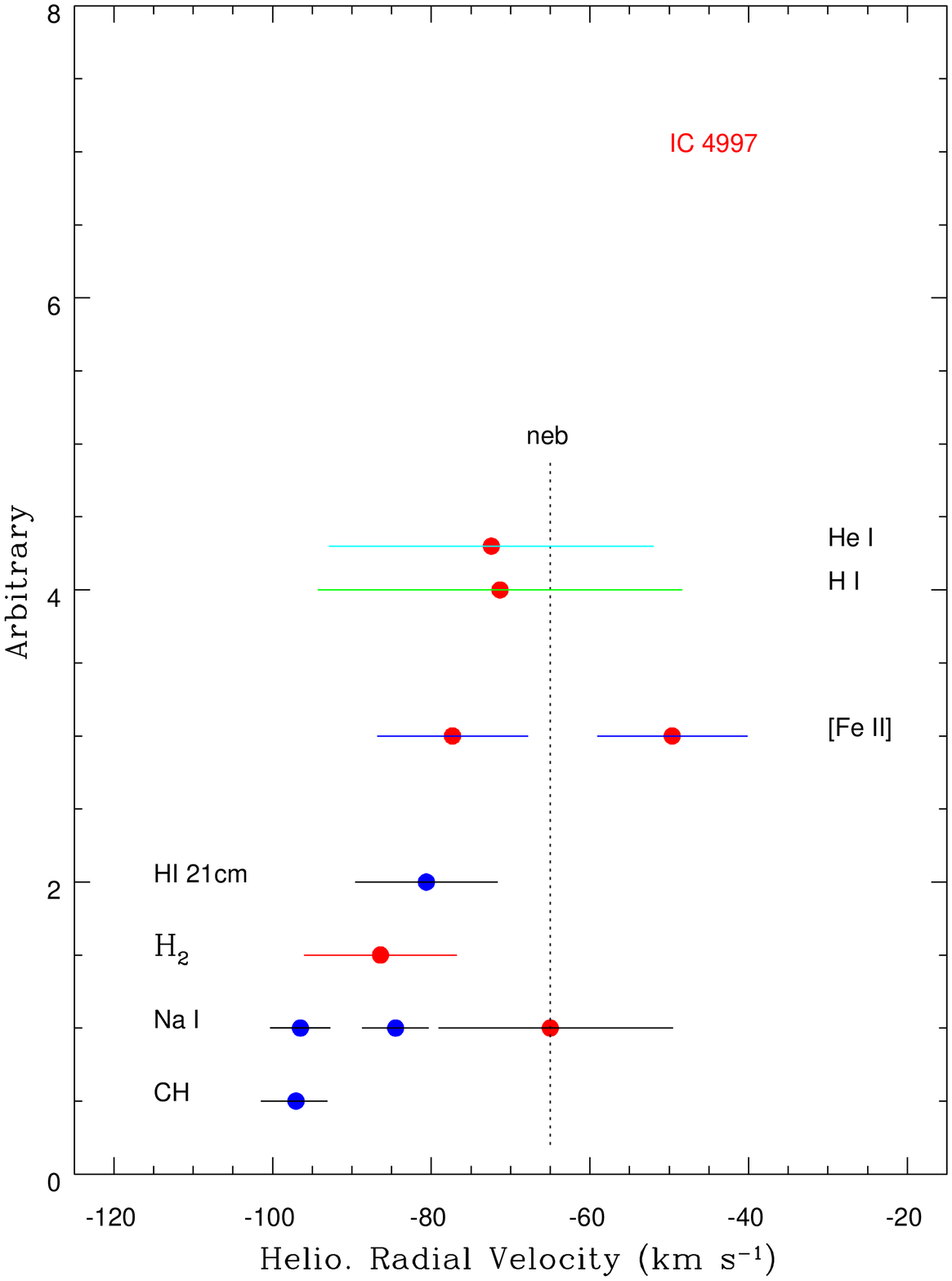}
\vspace{-0.8cm}
\caption{ Distribution of Radial Velocities of various species. The horizontal lines illustrates the average gaussian full width at half maximum (gfwhm) of lines of various species corrected for instrumental broadening.  The emission lines are denoted by red dots and absorptions by blue dots.  }
\end{figure*}

\section{The neutral envelope -- geometry}
Our exploration of IC\,4997's  neutral envelope concludes here with  discussion of the geometry and kinematics of the circumstellar environment. A simple model consists  of the hot rapidly evolving and mass-losing central star surrounded by expanding layers of ionized gas and an outer neutral envelope of gas and dust. Ionized gas around IC\,4997 is presumably of two possible origins: hot gas recently contributed by the central star's stellar wind and   gas (and dust) ejected by the star in its AGB phase  but now ionized either by ultraviolet photons from the central star or by interactions with the expanding shells of ionized gas.  The neutral envelope is considered  to be  the surviving remnant of the AGB star's circumstellar envelope.  An {\it Hubble Space Telescope}  image in H$\alpha$ (Sahai et al. 2011) shows that the most intense emission comes from an unresolved region around the central star but with faint `closed'  lobes extending out to about 1.4 arc secs from the central star. Our optical and infrared observations provide a seeing-averaged spectrum of this young PN with most light  provided by the central unresolved object.

With little information about the spatial structure contributing to the observed spectra, the focus is on the radial velocity information (Figure 6).  A datum of interest is the systemic velocity. Too little information is as yet available to determine the mean velocity of the central star. Emission lines provide a measure of the systemic velocity provided the gas is symmetrically placed about the star with respect to  emission strength and velocity. As noted above, the infrared He\,{\sc i} and H\,{\sc i} emission lines which are single-peaked with very nearly symmetrical profiles give mean velocities of about $-71$ km s$^{-1}$. Emission lines in the optical and infrared  from heavier ions appear double-peaked (Figures 2 and 3) with blue and red peaks often of different strength and with a central velocity near $-65$ km s$^{-1}$. Red and blue emission peaks are shifted each by about 15 km s$^{-1}$ with respect to the central velocity. Double-peaked profiles suggest emission arises from a shell which is rotating and/or rotating   or a bipolar flow. Such a bipolar flow, if present, is most likely not the flow identified by Sahai et al. (2011) from examination of {\it HST} images; the {\it HST}-identified bi-polarity is too faint to provide the prominent emission lines. (The small difference in the apparent systemic velocity of $-65$ km s$^{-1}$ and  $-70$ km s$^{-1}$ suggested by the H\,{\sc i} and He\,{\sc i} lines may  reflect temperature and density variations and optical depth effects.)

Residents of the neutral envelope are detected at velocities which are not characteristic   of the emission lines centered at the systemic velocity. 
The Na\,{\sc i}  D lines show  absorption components at $-84$ and  $-97$ km s$^{-1}$. The  CH$^{+}$ and CH absorption is at $-97$ km s$^{-1}$ with no observable absorption at $-84$ km s$^{-1}$.  For a systemic velocity of $-65$ km s$^{-1}$, these components imply expansion velocities of $-19$ and $-32$ km s$^{-1}$ of the neutral envelope. Such velocities are typical of expansion velocities for a circumstellar envelope of an AGB star. The absorption components (FWHM of 8 km s$^{-1}$) are barely resolved.  Evidently, the line of sight through the envelope to the central star is clumpy. 
  
The H$_2$ emission is attributed here to a shock created at the interface between the leading edge of the expanding region of ionized gas and neutral gas close to the inner edge of the neutral envelope.  If this interface extends around the central star, the H$_2$ emission is expected to occur at radial velocities spread over twice the expansion velocity relative to the systemic velocity, i.e., a profile not dissimilar to that of the typical double-peaked ionized line.  But, as Figure 3 clearly shows, the H$_2$ emission lines have a very different profile with emission centered seemingly on the $-84$ km s$^{-1}$ Na D component but with emission at the $-97$ km s$^{-1}$ Na D, CH and CH$^+$ component. The H$_2$ emission is spectrally resolved but does not include emission at the systemic velocity, i.e., the  emission appears to arise from a fairly wide arc centered roughly about the line of sight to the star.  This inferred geometry suggests that a bipolar flow is active along the line of sight.  Weak H$_2$ emission (Figure 3) is present at $+27$ km s$^{-1}$ or $43$ km s$^{-1}$ from the systemic velocity. This emission appears coincident with the red edge of infrared emission lines and may be linked to the interaction of the other half of the bipolar flow  into the neutral gas. Weakness of this  receding emission may be due to several factors: irregular distribution of material in the neutral envelope, obscuration by dust etc.  A spatial separation of H$_2$ emission from emission lines contributed by ionized gas has been seen also in other objects, for example, in the PN NGC\,7027 (Cox et al. 2002) and the shells around the post-AGB star IRAS 16594-4656 (Van de Steene et al. 2008).
 
\section{Concluding remarks}
High-resolution optical spectroscopy has provided the first detections of the CH and CH$^+$ molecules in absorption from a planetary nebula. The molecules most likely reside in remnants of the cool circumstellar shell of the AGB star which has  recently evolved into the hot central star of the
compact nebula IC\,4997. High-resolution infrared spectra across the H and K bands provide a rich emission line spectrum including several quadrupole vibration-rotation lines of the H$_2$ molecule. Comparison of emission line profiles suggest that the H$_2$ emission occurs where the ionized gas impacts  the cool circumstellar shell. Presence of H$_2$ molecules in regions likely to contain C$^+$ assists the formation of CH$^+$ through an endothermic reaction. It remains unclear why one but not the other of the two circumstellar components providing Na D in absorption contains CH and CH$^+$.   Additional insights into the present structure and past recent evolution of IC\,4997 may come by combining high spectral resolution in optical and infrared bands with high spatial resolution but this will be a challenge given the small angular size of IC\,4997.

\acknowledgements
{We thank the referee for carefully reading the manuscript and for giving a thoroughly helpful referee's report that helped improving the quality of the paper. This research has made use of the SIMBAD database, operated at CDS, Strasbourg, France. NKR would like to thank Department of Science and Technology (DST) for the support through the grant SERB/F/2143/2016-17 ''Aspects in Stellar and Galactic Evolution". NKR acknowledges financial support from the Spanish Ministry of Economy and Competitiveness (MINECO) under the 2015 Severo Ochoa Program MINECO SEV-2015-0548. 
NKR would also like to thank An\'{i}bal Garc\'{i}a-Hern\'{a}ndez (DAGH) and  Arturo Manchado (AM) for their kind hospitality during his visit to Tenerife and Tom Geballe for his comments on a draft of the paper. We also would like to thank Jorge Garc\'{i}a Rojas of IAC for his help with interpretation of the emission lines. 
DLL thanks the Robert A. Welch Foundation of  Houston, Texas for support through the grant F-634. 
DAGH and AM acknowledge support  from the State Research Agency (AEI) of the Spanish Ministry of Science, Innovation and Universities (MCIU)  and the European Regional Development Fund (FEDER) under grant AYA2017-88254-P.
JJDL acknowledges support provided by the MCIU grant AYA2016-78994-P.
This work is based on service observations made with the Nordic Optical Telescope operated on the island of La Palma by the Nordic Optical Telescope Scientific Association in the Spanish Observatorio del Roque de Los Muchachos of the Instituto de Astrof\'{i}sica de Canarias. This paper includes
data taken at The McDonald Observatory of The University of  Texas  at  Austin. We thanks Kyle Kaplan and Jacob McLane for help observing with IGRINS.
The Immersion Grating INfrared Spectrometer (IGRINS) was developed under a collaboration between the University of Texas at Austin and the Korea
Astronomy and Space Science Institute (KASI) with the financial support from the W.J. McDonald Observatory, the US National Science Foundation under
grant AST-1229522 to the University of Texas at Austin, and of the Korean GMT Project of KASI.  }


\end{document}